\documentclass{PoS}
\usepackage{amsmath,amssymb,amsfonts,macros,patricks}
\usepackage{wrapfig}
\title{$|V_{ub}|$ determination in lattice QCD }

\ShortTitle{$|V_{ub}|$ in lattice QCD}
\author{\includegraphics[width=2.5cm,angle=0]{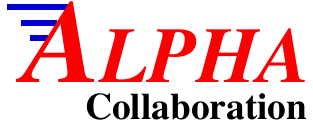}\hfill\parbox{3.5cm}{\footnotesize\it IFIC/12-77\\ CERN-PH-TH/2012-327\\ 
DESY 12-209\\ HU-EP-12/47 \\
SFB/CPP-12-87 
}}
\author{F.~Bahr$\,^{a}$, \speaker{F.~Bernardoni}$\,^{a}$, B.~Blossier$\,^{b}$,
J.~Bulava$\,^{c}$, M.~Della\ Morte$\,^{h}$, P.~Fritzsch$\,^{e}$,
N.~Garron$\,^{f}$, A.~G\'erardin$\,^{b}$, J.~Heitger$\,^{g}$,
G.~von~Hippel$\,^{d}$, A.~Ramos$\,^{a}$, H.~Simma$\,^{a}$, R.~Sommer$\,^{a}$\\
$^a$~NIC, DESY, Platanenallee~6, 15738~Zeuthen, Germany\\
$^b$~Laboratoire~de~Physique~Th\'eorique, CNRS/Universit\'e~Paris~XI,
F-91405~Orsay~Cedex, France\\
$^c$~CERN, Physics~Department, TH~Division, CH-1211~Geneva~23, Switzerland\\
$^d$~Institut~f\"ur~Kernphysik, University~of~Mainz, Becher-Weg~45, 55099~Mainz, Germany\\
$^e$~Institut~f\"ur~Physik, Humboldt-Universit\"at~zu~Berlin, Newtonstr.~15, 12489~Berlin, Germany\\
$^f$~School~of~Mathematics, Trinity~College, Dublin~2, Ireland\\
$^g$~Universit\"at~M\"unster, Institut~f\"ur~Theoretische~Physik, Wilhelm-Klemm-Str.~9, 48149~M\"unster, Germany\\
$^h$~IFIC, c/ Catedr\'atico Jos\'e Beltr\'an  2, E-46980 Paterna - Espa\~na
\email{fabio.bernardoni@desy.de}}

\abstract{The 2012 PDG reports a tension at the level of $3 \sigma$ between two exclusive determinations of $|V_{ub}|$. They are obtained by combining the experimental branching ratios of $B \to \tau \nu$ and $B \to \pi l \nu$ (respectively) with a theoretical computation of the hadronic matrix elements $\fB$ and the $B \to \pi$ form factor $f_+(q^2)$. To understand the tension, improved precision and a careful analysis of the systematics involved are necessary. We report the results of the ALPHA collaboration for $\fB$ from the lattice with 2 flavors of $O(a)$ improved Wilson fermions. We employ HQET, including $1/m_b$ corrections, with pion masses ranging down to $\approx$ 190 MeV. Renormalization and matching were performed non-perturbatively, and three lattice spacings reaching $a^{-1}\approx 4.1$ GeV are used in the continuum extrapolation. We also present progress towards a computation of $f_+(q^2)$, to directly compare two independent exclusive determinations of $|V_{ub}|$ with each other and with inclusive determinations. Additionally, we report on preliminary results for $\fBq{s}$, needed for the analysis of $B_s \to \mu^+\mu^-$.}

\FullConference{36th International Conference on High Energy Physics,\\
		July 4-11, 2012\\
		Melbourne, Australia}

\begin{document}

\section{Motivation}

The precise determination of the CKM matrix elements is a key for testing the Standard Model. 
Violations of CKM unitarity or discrepancies between independent determinations of the same matrix element can provide hints of New Physics.
At the time when we started our work, a tension at the level of $3\sigma$  
between two exclusive determinations of $|V_{\rm ub}|$ existed, as reported e.g. in the PDG of 2012. 
These determinations use the branching ratio (BR) for the processes $B\to \pi l 
\nu$ and $B\to \tau \nu$ from experiment combined with the form factor $f_+(q^2)$ 
and the B decay constant $f_{\rm B}$, respectively, from the lattice. 
Also an inclusive determination, based on a perturbative expansion in $\alpha_{\rm s}$ and an expansion in $1/m_b$, is possible \cite{Bigi:1993fe}. 
The results reported by the PDG 2012, computed before ICHEP 2012, can be summarized as follows \cite{Beringer:1900zz}:
\begin{eqnarray}
|V_{\rm ub}| &=& 0.00323(31) \qquad (B\to \pi l \nu)\,, \qquad
|V_{\rm ub}| = 0.00510(47) \qquad (B\to \tau \nu) \,, \\
|V_{\rm ub}| &=& 0.00441(34) \qquad \mbox{(inclusive)}\,\,. \nonumber 
\end{eqnarray}
At ICHEP 2012 the Belle collaboration reported a new result for BR$(B\to \tau \nu)$ \cite{Yook:2012} based on a new set of data, obtained with a more sophisticated tagging of the B. This result, taken alone, would yield a 
value for $|V_{\rm ub}|$ that is consistent with the exclusive determination from $B\to \pi$. However, more data and a careful inspection of the systematics involved are needed to draw more definitive conclusions.\\
While the experimental precision in the differential decay rate for $B \to \pi l \nu$ has by now reached good precision, $B \to \tau \nu$ events are more difficult to reconstruct and there is an error of the order of $20 \%$ on the branching ratio. The situation on the theoretical side is the opposite: the lattice computation of a form factor is more challenging 
than that of a decay constant.
\\
At this conference we have presented the results for the determination of $f_{\rm B}$ by the ALPHA collaboration which use fully non-perturbative 
renormalization and matching, and CLS configurations with two degenerate dynamical quarks in the sea.  A parallel effort to determine $f_+(q^2)$ in the same setup is ongoing: we have presented the progress reached so far, and the precision that we expect to achieve, once the full non-perturbative renormalization and matching at order $1/m_b$ have been completed. The comparison of these two exclusive predictions, in which the relevant hadronic parameters have been computed in the same setup, will provide a test  as free as possible from systematics.\\
Recently LHCb has presented the first evidence for $B_s \to \mu^+\mu^-$, with a decay rate compatible with the Standard Model \cite{Bstomumu}.
We have presented the determination by the ALPHA collaboration of the $B_s$ decay constant, $f_{\rm B_s}$, that enters in the theoretical prediction of this decay.

\section{HQET}

Our computations were performed on CLS configurations, which have two degenerate $O(a)$ improved Wilson quarks. The ensembles used in this work have pion masses $m_\pi$ in the range $190\, {\rm MeV} \lesssim m_\pi \lesssim 450\, {\rm MeV}$ at three lattice spacings $a$, namely $a \in \{ 0.078,\, 0.065,\, 0.045\} {\rm fm}$. All of them have a spatial extent $L$ such that $m_\pi L>4$, so that volume effects are expected to be very small.  \\
Even for our finest lattice spacing the $b$ quark cannot be simulated directly, given that $am_b > 1$. However, for the low energy processes we are interested in, the $m_b$ scale can be integrated out. Our approach is to use HQET, which is an expansion of the QCD Lagrangian in powers of $1/m_b$. At leading order the $b$ quark is static, i.e. the Lagrangian involves no space derivatives. If we include terms up to order $1/m_b$ the Lagrangian becomes:
\be
\lag{HQET}(x)
=
\;\displaystyle\heavyb(x)\,D_0\,\heavy(x)
\;-\,{\omkin}\Okin (x) -{\omspin}\Ospin(x)\,,
\ee\be
\Okin(x)  = \heavyb(x)\,\vecD^2\,\heavy(x)\,,
~~~~~~~ \Ospin(x) = \heavyb(x)\,\vecsigma\cdot\vecB\,\heavy(x).\nonumber
\ee
The corresponding expansion of the time component $A_0$ of the
    heavy-light current  (at $p=0$) is
\bea
\Arenhqet= 
{\zahqet}\,\Big[\,\Astat+{\cah{1}}\Ah{1}\,\Big]\;\;,\;\; 
\Astat=
\lightb\,\gamma_0\gamma_5\,\heavy\;\;,\;\;
\Ah{1}= 
\lightb\,\gamma_5\gamma_i\,\half\,(\nabsym{i}\,-\!\lnabsym{i}\,)\,\heavy.
\eea
Since in HQET the $\rmO(1/m_b)$ terms appear only as insertions in
   correlation functions, HQET is renormalizable order by order in
   $1/m_b$ because the static theory is. Once the HQET parameters 
   $\omega_i \in \{ m_{\rm bare},\zahqet,\cah{1},\omega_{\rm kin},\omega_{\rm spin} \}$
    have been determined using {\em non-perturbative} matching
   \cite{Heitger:2003nj}, the continuum limit can therefore be taken safely.

\section{Matching}

The matching was performed in the Schr\"odinger Functional scheme
in a small volume $L_1\approx 0.4$ fm, where $a\mb\ll 1$, and relativistic b-quarks
   can be simulated \cite{Blossier:2012qu}. The HQET parameters can then
   be fixed by imposing the matching conditions for suitable observables $\Phi$
\bea
\PhiHQET(L_1,M,a)=\PhiQCD(L_1,M,0)\,,~~~~~
\PhiQCD(L_1,M,0)=\lim_{a\to 0}\PhiQCD(L_1,M,a)\,,
\eea
and hence the $\omega_i$ inherit their dependence on the heavy-quark
   mass $M$ from QCD. In particular here $M$ is the RGI mass of the b-quark \cite{Fritzsch:2010aw}. Using
   finite-size scaling recursively, we can then take the step $L_1\to L_2=2L_1$, and
   finally connect with large volumes $L_{\infty}\gtrsim \max(2~\Fm,4/m_\pi)$.

Having performed this matching procedure for each of the lattice
   spacings used in our large-volume simulations, we know the
   corresponding $\Nf=2$ HQET parameters $\omega_i(L_1,M,a)$
   non-\hyphenation{per-tur-ba-ti-ve-ly} for a range of values of $M$ in the neighbourhood
   of the $b$ quark mass.

\section{Results} 

The HQET energies and matrix elements are extracted at large Euclidean time $t$ separations. The effects of excited states are exponentially suppressed like $\sim e^{-(E_2-E_1)t}$, where $E_2$ and $E_1$ are the energies of the first  excited state and the ground state, respectively. To achieve a better suppression, we solve the Generalized Eigenvalue Problem
(GEVP)~\cite{Blossier:2009kd} 
\begin{align}
  C(t)v_n(t,t_0) = \lambda_n(t,t_0)C(t_0)v_n(t,t_0)\,, ~~~~~~~~~ t_0<t<2t_0\,,
\end{align}
for an $N\times N$ correlator matrix $C(t)$ with $N=3$. Each entry of
the matrix corresponds to a different Gaussian smearing
level of the light quark field in the B-meson interpolating quark bilinear. 
The corrections to the energies and matrix elements so obtained behave like 
$\propto \exp\left\{ -(E_{N+1}-E_1)t\right\}$ and  $\propto \exp\left\{ -(E_{N+1}-E_1)t_0\right\}
\times\exp\left\{ -(E_2-E_1)(t-t_0)\right\}$, respectively. The residual systematic errors are kept under control by requiring $\sigma_{\rm stat}\gtrsim 3\sigma_{\rm sys}$.

In phenomenological predictions, we need to know the HQET parameters at the physical mass of the $b$ quark  $\omega_i(L_1,M_{\rm b},a)$. To this end we impose  $\mB(L_1,M_{\rm b},m_{\pi}^{\rm exp},a=0) \equiv \mB^{\rm exp} = 5279.5\, \MeV$. The mass of the $B$ at physical pion mass $m_{\pi}^{\rm exp}$ is obtained through a chiral and continuum extrapolation \cite{Bernardoni:2009sx}:
\begin{align}
 \mB\left(z,m_\pi,a,{\rm n}\right) &= B(z)+ C m^2_\pi - \frac{3\widehat{g}^2}{16\pi f_\pi^2} m^3_\pi  + D_{\rm n} a^2 ,
&  \widehat{g} =0.51(2)\,\cite{Bulava:2010ej}.
\end{align}
In HQET the B-meson mass is given by  
$\mB ={ m_{\rm bare}} + E^{\rm stat} + { \omega_{\rm kin}}  E^{\rm kin}+ { \omega_{\rm spin}} E^{\rm spin}$, so that
\begin{align}
   M_{\rm b} &= 6.56(15)(06)_{z}\,\GeV \;, & &\text{or equivalently} &
   \overline{m}_{\rm b}^{\overline{MS}}(\overline{m}_{\rm b}) &= 4.22(10)(4)_{z}\,\GeV  \;. 
\end{align}
For the following analyses, we use the values of HQET parameters
   obtained from an interpolation to $z\equiv z_{\rm b}$.
\\

\subsection{$B$ and $B_s$ decay constants}

\begin{figure}
\begin{center}
\includegraphics[width=\textwidth,keepaspectratio]{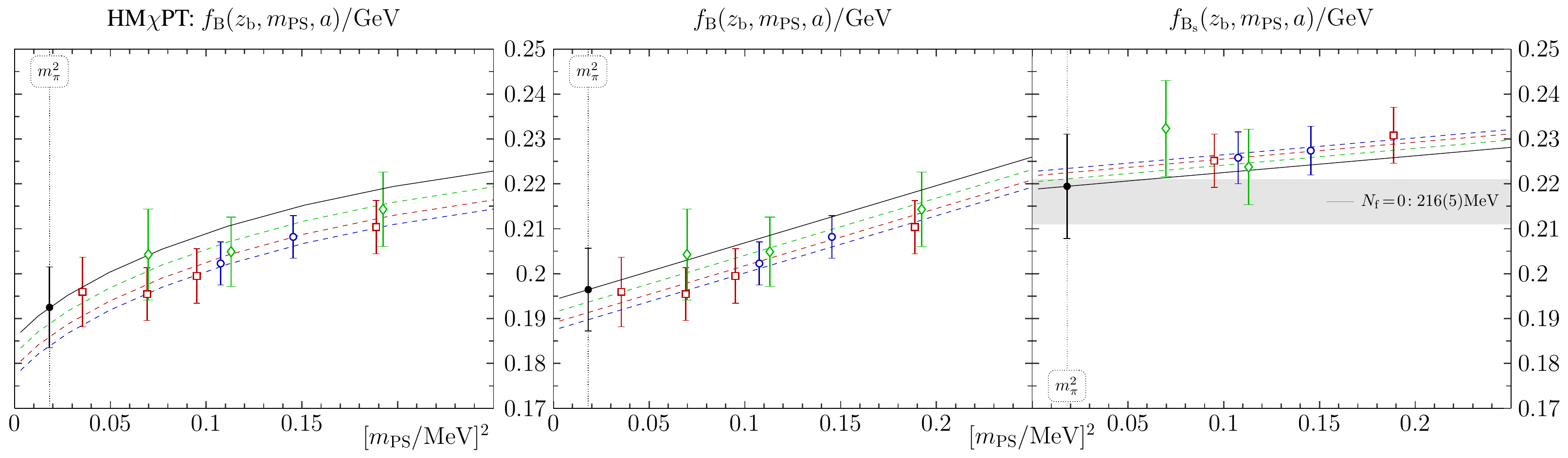}
\end{center}
\caption{{\em Left:} HM$\chi$PT extrapolation of $\fB$;
         {\em centre/right:} linear extrapolation of $\fB$ and $\fBs$.
         The blue, red and green points correspond to ensembles at
         $a=0.075$~fm, $0.065$~fm and $0.048$~fm, respectively.
	 The fit formulae evaluated at each given lattice spacing are
         shown in colour. The black curve is the chiral and continuum 
	 extrapolation. It is a fit to all the points shown in the figure plus 
	 other at the same pion masses but obtained using a different discretization 
	 for the heavy quark action. }
\label{extrap}	 
\end{figure}

Using the HQET formulas to combine the matrix elements and energies with the matching parameters at $z\equiv z_{\rm b}$ we can compute $\fB$ and $\fBq{s}$ through a chiral and continuum extrapolation dictated by HMChPT \cite{Sharpe:1995qp,Goity:1992tp}:
\be
 \fB\left(m_{\pi},a,{\rm n}\right) = b\left[1-\frac{3}{4} \frac{1+3\widehat{g}^2}{(4\pi f_\pi)^2}
          m^2_{\pi}\ln (m^2_{\pi})\right] + c m^2_{\pi}   + d_{\rm n} a^2 \,. 
\ee
Our analysis for ICHEP 2012 gives
\bean
     \fB =  193(9)_{\rm stat}(4)_{\chi}\, \MeV\,, ~~~~~~~
 \fBq{s} =  219(12)_{\rm stat}\, \MeV \,,
\eean
where the error coming from the chiral extrapolation is determined by comparing to a linear extrapolation in $m_\pi$ (see Fig.~\ref{extrap}). For $\fBq{s}$ not all ensembles are
   analysed yet. For more details, see \cite{Bernardoni:2012ti}. Our value for $\fB$ is compatible with the values
   found by other collaborations \cite{Na:2012kp,Bazavov:2011aa,McNeile:2011ng,Dimopoulos:2011gx}.

\subsection{$B \to  \pi$ form factor}

The form factor $f_+(q^2)$ is defined through the Lorentz decomposition of the matrix element:
\beq
\langle \pi(p_\pi) | V^\mu | B(p_{\rm B} ) \rangle = f_+(q^2)\left[ p_{\rm B}^\mu + p_\pi^\mu-\frac{m_{\rm B}^2-m_\pi^2}{q^2}q^\mu \right]+f_0(q^2)\frac{m_{\rm B}^2-m_\pi^2}{q^2}q^\mu \,,
\eeq
where $q^\mu=p_{\rm B}^\mu-p_\pi^\mu$. To extract this matrix element from lattice simulations, we consider the ratio:
\bea
\label{ratio}
&&R(t_\pi,t_{\rm B})\equiv\frac{ 
\sum\limits _{\vec{x}_\pi, \vec{x}_{\rm B}}e^{-i\vec{p}  \cdot \vec{x}_\pi}
\langle   P_{ll}(t_\pi+t_{\rm B},\vec{x}_\pi)  V^\mu (t_{\rm B},\vec{x_{\rm B}})  P_{hl}(0)  \rangle}
{\sqrt{\sum\limits_{\vec{x}_\pi}e^{-i\vec{p}  \cdot \vec{x}_\pi}\langle  P_{ll}(x_\pi) P_{ll}(0)  \rangle  \sum\limits_{\vec{x}_{\rm B}} \langle  P_{hl}(x_{\rm B})   P_{hl}(0)  \rangle}}\,, \nonumber \\
&&\langle \pi(p_\pi) | V^\mu | B(p_{\rm B} ) \rangle =  \lim_{T,\,t_{\rm B},\,t_\pi\to \infty} R(t_\pi,t_{\rm B})e^{E_\pi t_\pi/2+m_{\rm B} t_{\rm B}/2}
\eea
where $P_{ll}$ and $ P_{hl}$ are interpolating operators for the $\pi$ and the $B$ meson, respectively.
\begin{wrapfigure}{r}{0.5 \textwidth}
\begin{center}
\includegraphics[width=11cm]{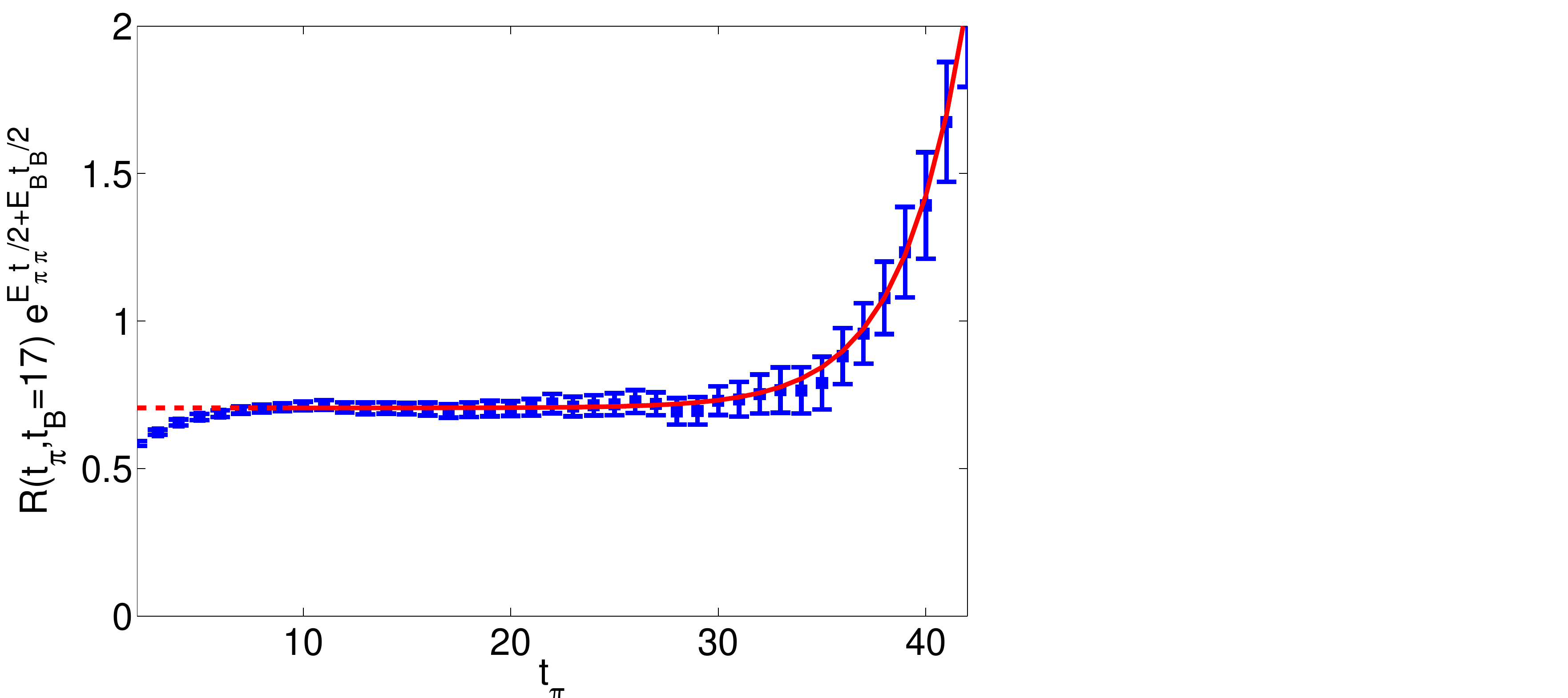} 
\end{center}
\caption{$t_\pi$ dependence for the considered ratio eq.~\protect \eqref{ratio} on one of our lattices with $m_\pi \approx 300$ MeV. We use a smeared interpolating operator for the B and the $\pi$. The red curve is obtained by fitting the finite $T$ effects with the prediction from transfer matrix theory:
$
R(t_\pi,t_B)=\frac{
 A+B_{t_{\rm B}} e^{-E_\pi(T-2t_\pi)}}
 {\sqrt{1+e^{-E_\pi (T-2t_\pi)}}}
$
with $A$ and $B$ as parameters.}
\label{smearing}
\end{wrapfigure}
One additional difficulty in comparison to the extraction of decay constants is the presence of large finite $T$ effects, where $T$ is the temporal extension of the lattice. So far we have restricted to few lattices and to the static limit to demonstrate the feasibility of this computation in our setup. Our results show that finite $T$ effects can be understood in the transfer matrix formalism (see Fig.~\ref{smearing}) and that a precision of $5\%$ is achievable. \\
Combining our result at largest $q^2$, $q=p_\pi-p_B$, with the experimental data available and a parametrization of the $q^2$ dependence based on very general properties like unitarity and analyticity \cite{Bourrely:2008za}, $|V_{ub}|$ is obtained with a $15\%$ precision. This does not include the necessary extrapolation in the light quark mass and the lattice spacing. These missing steps will soon be coming out. For more details, see \cite{Bahr:2012vt}.

\section*{Acknowledgments}

\small

This work is supported in part
by the SFB/TR~9
and grant HE~4517/2-1 (P.F. and J.H.)
of the Deutsche Forschungsgemeinschaft
and by the European Community through
EU Contract MRTN-CT-2006-035482, ``FLAVIAnet''.
We thank our colleagues in the CLS effort for the joint production
and use of gauge configurations.
We gratefully acknowledge the computer resources provided
within the Distributed European Computing Initiative
by the PRACE-2IP,
with funding from the European Community's Seventh Framework Programme
(FP7/2007-2013) under grant agreement RI-283493,
by the Grand \'Equipement National de Calcul Intensif at CINES in Montpellier,
and by the John von Neumann Institute for Computing
at FZ~J\"ulich, at the HLRN in Berlin, and at DESY, Zeuthen.


\small

\bibliographystyle{ieeetr}
\bibliography{lattice}

\end{document}